\documentclass[twocolumn,showpacs,preprintnumbers,pre,fleqn]{revtex4}
\usepackage{epsfig}
\usepackage{graphicx}
\usepackage{amsfonts}

\begin{document}

\title{Correlated electrons systems on the Apollonian network}
\author{Andre M. C. Souza$^{1,2}$ and Hans Herrmann$^{3,4}$}

\affiliation{$^{1}$Institut f\"{u}r Computerphysik, Universit\"{a}t Stuttgart, Pfaffenwaldring 27, 70569 Stuttgart, Germany}

\affiliation{$^{2}$Departamento de Fisica, Universidade Federal de Sergipe, 49100-000 Sao Cristovao-SE, Brazil}

\affiliation{$^{3}$Institut f\"{u}r Baustoffe, ETH H\"{o}nggerberg, HIF E 12, CH-8093 Z\"{u}rich, Switzerland}

\affiliation{$^{4}$Departamento de F\'{i}sica, Universidade Federal do Cear\'{a}, 60451-970 Fortaleza-CE, Brazil}

\date{\today}
\begin{abstract}
Strongly correlated electrons on an Apollonian network are studied using 
the Hubbard model. Ground-state and thermodynamic properties,
including specific heat, magnetic susceptibility, spin-spin correlation function,
double occupancy and one-electron transfer, are evaluated applying direct diagonalization
and quantum Monte Carlo. The results support several types of magnetic behavior.
In the strong-coupling limit, the quantum anisotropic
spin $\frac{1}{2}$ Heisenberg model is used and the phase diagram is discussed
using the renormalization group method.
For ferromagnetic coupling, we always observe the existence of 
long-range order. For antiferromagnetic coupling, we find
a paramagnetic phase for all finite temperatures.
\end{abstract}
\maketitle
\def\trazo{\mathop{\rm Tr}\rm_{site \; 4}}

\section{Introduction}
Scale-invariant networks have been subject of
intensive study in view of possible insights into inhomogeneous problems
such as random magnets, surfaces, porous rocks, aerogels, sponges, etc.
They are relevant to many different real complex situations like biological,
social and technological systems \cite{net,mendes},
where the structures can embody a particular graph, e.g.,
a scale free network. In this case, the fraction of sites
with $k$ connections follows a power law \cite{BA}. This 
topology has been found, in particular, on World Wide Web and Internet
networks \cite{fa}. 

More recently, the area of scale-invariant networks has been
highly motivated by the creation of a synthetic nanometer-scale Sierpinski hexagonal gasket,
a self-similar fractal macromolecule \cite{kome}. New types of photoelectric cells, molecular batteries and energy storage may be possible. This perspective is relevant in view of the renewed interest in correlated electron systems on these networks, like studies of quantum magnetism, superconductivity, metal-insulator transition, etc. An important aspect that can be analyzed is the influence of the topology and the appearance of magnetic order. 

In this paper, we investigate one family of free scale-invariant graphs,
the Apollonian network \cite{hans1,hans2,ap2}.
The purpose is to examine correlated electrons on Apollonian networks.
We explore the connection between the scale-invariant network topology
and the magnetic properties of quantum magnetism models, namely, Hubbard and 
anisotropic spin $\frac{1}{2}$ Heisenberg models.
We have examined the Hubbard model using the small cluster diagonalization \cite{cal,meuDD} 
and quantum Monte Carlo \cite{Hir,meuMC}. The Heisenberg model is studied
within a real-space renormalization group framework \cite{car,mar,meu}. 

The construction of the Apollonian network has the topology of the contacts of an
Apollonian packing of two-dimensional disks. As initial configuration
we use, for an Apollonian packing, three mutually touching circles inscribed
inside a circular space.
The interstices of the initial disks define a curvilinear triangle
to be filled. In the next iteration, four disks are inscribed, each touching all the
sides of the corresponding curvilinear triangle. This process is repeated indefinitely 
by setting disks in the newly generated curvilinear triangles.
The Apollonian packing is a fractal whose dimension has been estimated 
as 1.3057 \cite{ap2}.

\begin{figure}[!ht]
\psfig{figure=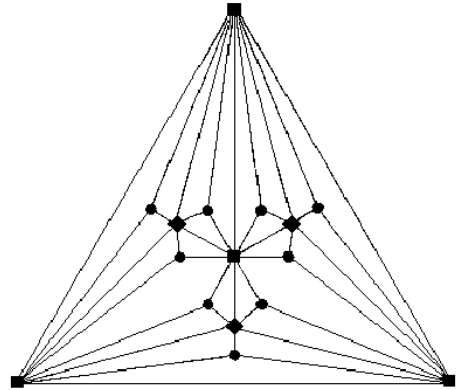,width=70mm,angle=0}
\caption{First three generations of the Apollonian network. Sites represented by squares, diamonds and circles are introduced in the first, second and third generation.}
\end{figure}

The Apollonian network (AN) is derived from the two-dimensional Apollonian packing 
associating vertices with the circles and
connecting two vertices if the corresponding circles touch. Fig 1 shows
the first three generations of the AN. With this construction
procedure one obtains a deterministic scale free network
where the number of sites at iteration $n$ is $(3^{n-1}+5)/2$.
Besides being a scale-free network, the AN has interesting properties like
being Euclidean, matching and space-filling \cite{hans1}.

The organization of this paper is as follows.
The models are briefly introduced in section II. 
The results for ground-state and thermodynamic properties 
of the Hubbard model on AN are presented in section III and IV,
respectively. The phase diagram of the quantum anisotropic Heisenberg
model is discussed in section V and conclusions are presented in section VI.

\section{Models}
Our study of the magnetic properties on AN are based on the Hubbard model.
The Hamiltonian of the Hubbard model is defined by
\begin{equation}
{\cal H}_{u}= -t \sum_{<ij>\alpha} c_{i\alpha}^{\dag}c_{j\alpha} +
U \sum_{i} n_{i\uparrow}n_{i\downarrow} ,
\label{hamil} 
\end{equation}
where $c_{i\alpha}^{\dag}$, $c_{i\alpha}$ and
$n_{i\alpha}\equiv c_{i\alpha}^{\dag}c_{i\alpha}$ are respectively the
creation, anihilation and number operators
for an electron with spin $\alpha$ in an orbital localized at site $i$; 
the $<ij>$ sum runs over nearest-neighbor sites on AN \cite{and}.

The question of magnetic order in the one-band Hubbard model has
been investigated by several authors and much controversy has arisen.
In the strong-coupling limit a major part of the ferromagnetic phase is predicted.
In this case, at half filling band, $U$ is much larger than $t$, 
and the Hubbard model, using a suitable expansion in perturbation theory,
is formally equivalent to the antiferromagnetic Heisenberg model. 
The Heisenberg exchange parameter $J$ is written in terms of the Hubbard model
parameters as $J = - 4t^2 /U$, where the Hamiltonian of the Heisenberg model is defined by
\begin{equation}
 {\cal H}_{e} = -J \sum_{<ij>} \vec{S}_{i} \vec{S}_{j},
\end{equation}
where $\vec{S}_{i}$ is the total spin $\frac{1}{2}$ operator for the $i$th site.
Here we study the anisotropic Heisenberg model to compare the behavior for different
values of the anisotropy parameter \cite{meu}. 

\section{Ground-state properties}
We have obtained exact numerical results of the Hubbard model defined
on second generation of the AN, corresponding to seven sites.
We computed all the eigenvalues and eigenvectors of the Hamiltonian of Eq. 1
on a basis of states for which the occupation number is diagonal. We consider subspaces
of fixed total azimuthal spin operator $S_Z$. Thus, the maximum possible dimension
 of the matrix to diagonalize is $1225$. 

\begin{table} 
\begin{tabular}{cl}
\cline{1-2} 
$n=2$ & $S=0$, all $U/t$ \\ 
$n=3$ & $S=1/2$, all $U/t$\\ 
$n=4$ & $S=1$, all $U/t$ \\ 
$n=5$ & $S=1/2$, all $U/t$ \\
$n=6$ & $S=0$, all $U/t$ \\  
$n=7$ & $S=1/2$, all $U/t$ \\ 
$n=8$ & $S=0$, $U/t < 12.5$ \\ 
      & $S=1$, $ 12.5 < U/t < 15.4$ \\ 
      & $S=2$, $ 15.4 < U/t < 20.1$ \\ 
      & $S=3$, $ U/t > y$ \\ 
$n=9$ & $S=1/2$, all $U/t$ \\ 
$n=10$ & $S=1$, $U/t < 8.44$ \\
       & $S=2$, $ U/t > 8.44$ \\ 
$n=11$ & $S=1/2$, $U/t < 0.44$ \\
       & $S=3/2$,   $U/t > 0.44$ \\
$n=12$ & $S=0$,  $U/t < 0.57$ \\
       & $S=1$, $U/t > 0.57$ \\

 \cline{1-2} 
\end{tabular}
\caption{Spin of the ground state as function of occupation number and $U/t$ on the
AN of 7 sites.}
\label{tab}
\end{table}

\begin{figure}
\psfig{figure=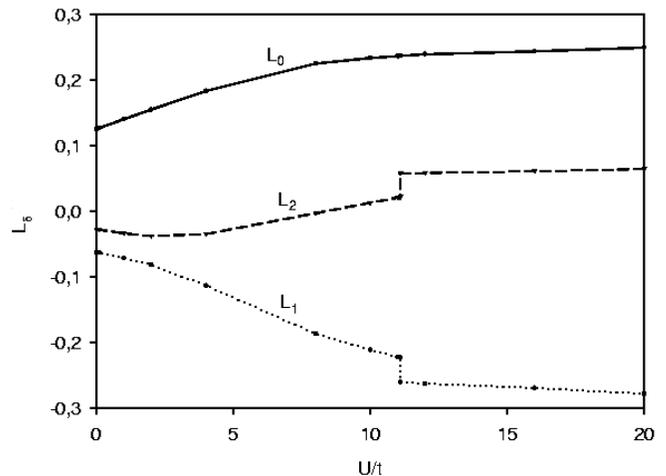,width=90mm,angle=0}
\caption{Ground state spin-spin correlation function versus $U/t$ for the half-filled band
on the AN of 7 sites.}
\end{figure}

Table I shows the results for the spin of the ground state. 
Until the half-filled band, the ocurrence of a ferromagnetic state, 
where the total spin $S$ is not minimum,
can be found only for four electrons. In this case the ground state is threefold-degenerate corresponding to a triplet state of $S=1$. This behavior can be explained considering
the spectrum of the free electron system. Table II lists the single electron energies and degeneracies. The double degeneracy in second level forms a triplet in
lowest energy for the four electrons case.
For two, three, five, six and seven electrons the ground state always has minimum spin.
We observe that the Coulombian interaction does not favour the occurrence of a high 
spin ground state. If $U/t$ is strong, the jumps of electrons decrease and a frustrated
ordered antiferromagnet is favored. 
We easily see that the network is not bipartite, so its structure is antiferomagnetically
frustrated. We find a similar competition, between interaction and frustration, 
in the low-temperature antiferromagnetic state on a triangular lattice \cite{tri}. 

\begin{table} 
\begin{tabular}{c|c|c|c|c}

$E_{1}$    & $E_{2}=E_{3}$ & $E_{4}$ & $E_{5}=E_{6}$ & $E_{7}$ \\ 
\cline{1-5} 
$-4.5114$  & $-0.6180$ & $0.7589$& $1.6180$ &  $1.7525$ \\ 

\end{tabular}
\caption{Energy levels $E_{i}$ (i=1, 2,..., 7) of free electron system ($U/t=0$)
 on the AN of 7 sites.}
\label{tab2}
\end{table}

Above the half-filled band we obtain a ferromagnetic ground-state.
Nagaoka demonstrated that the ferromagnetism is expected for
the antiferromagnetically frustrated structures in the half-filled band
case with one excess electron and $U \rightarrow \infty$ \cite{nag}. 
Here, this limit represents the eight electrons case for $U/t >> 1$. We can see that
if $U/t$ increases the alignment of spins also increases, in such a manner that, 
for $U/t > 20.1$, the ferromagnetic state with the maximum total spin is the ground state. 
For more than eight electrons ferrmagnetism is possible, but not stronger than
in the case of completely saturated ferromagnetism.

In general, the behavior of short(or long)-range ordering can be
better observed studying the spin-spin correlation function. 
Considering the azimuthal spin operator on site $i$ as 
$S^{z}_i=n_{i\uparrow}-n_{i\downarrow}$, the spin-spin correlation
function is defined, for $N$ sites, as
\begin{equation}
L_\delta =\frac 1{4N}\sum_i S^{z}_i S^{z}_{i+\delta } .
\end{equation}
The quantity $L_{0}$, called local moment, depends on the magnitude of the difference between
up and down electron spin at each site. It shows the degree of localization of
electrons. For a completely localized system in which each site is
occupied by a single electron (up or down), $L_{0}=1/4$, while for a
non-interacting system $L_{0}=1/8$. 
$L_{\delta }$ ($\delta \neq 0$) is the correlation between the electron spins at different sites. 
They are related to the magnetic ordering. 
The results for the ground state spin-spin correlation function versus $U/t$ in the half-filled
band case are presented in Fig. 2. $L_0$ gradually increases if $U/t>0$ increases, i.e.,
electrons gradually localize. $L_1$ has negative sign and increases if $U/t>0$ increases
and $L_2$ has a critical value for $U/t=8$ where the behavior changes from negative to
a positive sign.  For $U/t > 8$, $L_2$ is positive, $L_1$ is negative, and an
antiferromagnetic order appear. For $U/t=11.1$ there is a change in the spatial symmetry of the ground state. It is important to observe the discontinuous change of $L_1$ and $L_2$. 
This quantum transition point at $U/t=11.1$  must not be a Mott metal-insulator
 transition because $L_0$ is continuous \cite{cal}. For $U/t >> 1$, 
$L_{0}$ is approximately equal to $1/4$ and we can conclude that each site is occupied by
just one electron. Our results indicate that the system can present a 
ferrimagnetic order, however, because the size of the network considered, 
the ferrimagnetic state has minimum spin $S=1/2$, and we cannot have high spin ground state.  
For $U/t < 8$, $L_2$ is negative and the magnetic behavior of the ground state is more complex. 

We analyse the wave function of the ground state, defining $P_k=|\Psi _{k0}|^{2}$, where 
$|\Psi _{k0}|$ is the $n$th component of the ground eigenstate. We define the
total antiferromagnetic configuration ($k=AF$) such that sites introduced
in the first generation on AN (represented by squares in Fig. 1) are occupied by electrons 
of a type of spin and sites introduced in second generation (represented by diamonds in Fig. 1)
are occupied by electrons of opposite spin.  
We observe that $P_{AF}/P_{k} >2$ for $U/t=0$ and increases to  $P_{AF}/P_{k} > 15$ if
$U/t=8$, for all $k \ne AF$, i.e., the total antiferromagnetic configuration $P_{AF}$ is approximately $2$ times for $U/t=0$ and $15$ times for $U/t=8$, more probable than any
another configuration. Thus, we can construct a physical picture where the ground state has an antiferromagnetic order (frustrated) at any $U/t$.

\section{Thermodynamic properties}
On one hand, we study the thermodynamic properties of the half-filled band
Hubbard model using the method of small-cluster
exact-diagonalization calculations in the grand canonical ensemble \cite{meuDD}. 
We have calculated all eigenvalues and eigenfunctions for AN of 7 sites.
On the other hand, we study lattices of 63 and 124 sites using the grand canonical
quantum Monte Carlo method \cite{Hir,meuMC}.
We used a discrete Hubbard-Stratonovich transformation to
convert the problem into one of free particles interacting with a time-dependent
Ising field, together with an exact updating algorithm for the fermions Green's function to compute the relative weights of the Ising configurations. 

We consider several values of $U/t$ for all the studied
thermodynamic properties. We determine with considerable accuracy the
temperature dependence of the spin-spin correlation functions, spin susceptibility,
specific heat, double occupancy and one electron transfer.

\begin{figure}
\psfig{figure=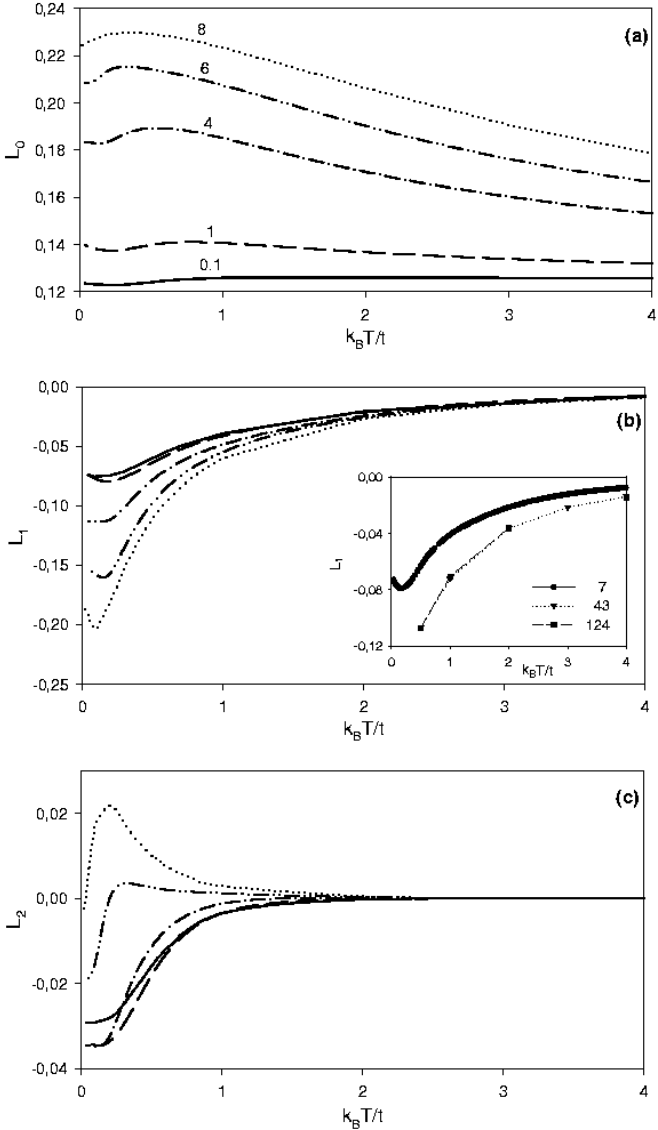,width=90mm,angle=0}
\caption{Correlation functions $L_\delta $ vs. temperature with $U/t=0.1$, 1, 6 and 8 for the AN of 7 sites. (a) $L_0$; (b) $L_1$; (c) $L_2$. In the inset of (b), $L_1$ with $U/t=1$ for
7, 43 and 124 sites on AN using Monte Carlo method.}
\end{figure}

The temperature dependence of $L_0$, $L_1$ and $L_2$ for some
typical values of $U/t$ is shown in Figs. 3a-3c. 
If $U/t$ increases, $L_0$ gradually increases for $U/t>0$, 
indicating that electrons are gradually localizing. 
At high temperature $L_0$ gradually decreases and electrons gradually delocalize. 
At very low temperatures, the temperature dependence with negative signs of $L_1$ induces
an antiferromagetic ordering.
The inset in Fig. 3b shows Monte Carlo results for AN with 7, 43 and 124 sites and $U/t=1$.
Results for 43 and 124 sites do not reveal any new behavior, 
however, the antiferromagetic ordering increases.
We observe that the properties obtained for 7 sites must be qualitatively
equivalent to those for larger AN.
We observe a competition between interaction and frustration in the ordered
antiferromagnetic state by the analysis of $L_2$ . For small or intermediary $U/t$,
the negative sign of $L_2$ is a result of the frustrated structure of AN, disfavoring
the antiferromagnetic order. For $U/t >> 1$, $L_2$ has positive sign, showing that the
strong electron interaction favors the antiferromagnetic order. 

\begin{figure}
\psfig{figure=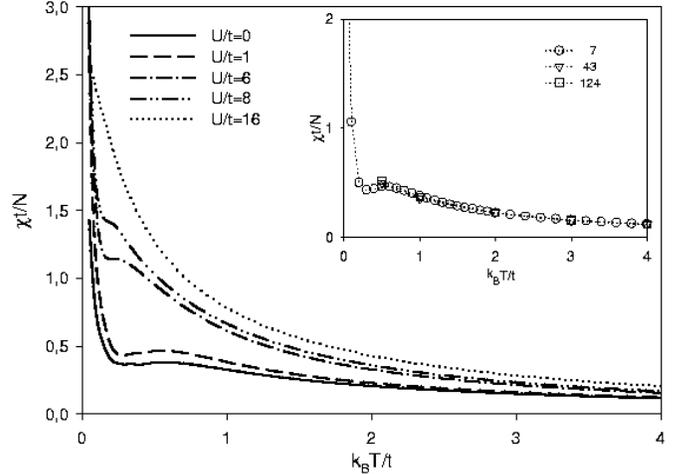,width=90mm,angle=0}
\caption{Magnetic susceptibility as a function of temperature with $U/t=0$, $0.5$, $4$, and $8$
for the AN of 7 sites. Inset: Magnetic susceptibility with $U/t=1$ for
7, 43 and 124 sites on AN using Monte Carlo method.}
\end{figure}

Fig. 4 shows the temperature dependence of the magnetic susceptibility, $\chi $. 
The Curie-Weiss behavior, $\chi \propto 1/(T-\theta )$, is observed for high temperatures.
For small and intermediary $U/t$ the curves present a peak. The temperature associated
 with the peak of $\chi $ corresponds approximately to the rapid decay
of $L_1$ and $L_2$, and is a consequence of the collective excitations that
lead to the destruction of the antiferromagetic order \cite{cal}.
Here the ground state has spin $S=1/2$ and $\chi$ must
go to infinity at temperature $T=0$. The susceptibility $\chi $ with $U/t=1$ for
7, 43 and 124 sites on AN using Monte Carlo method is shown in the inset of Fig. 4.
In this case, we do not see any discrepancy between the results for different clusters.

\begin{figure}
\psfig{figure=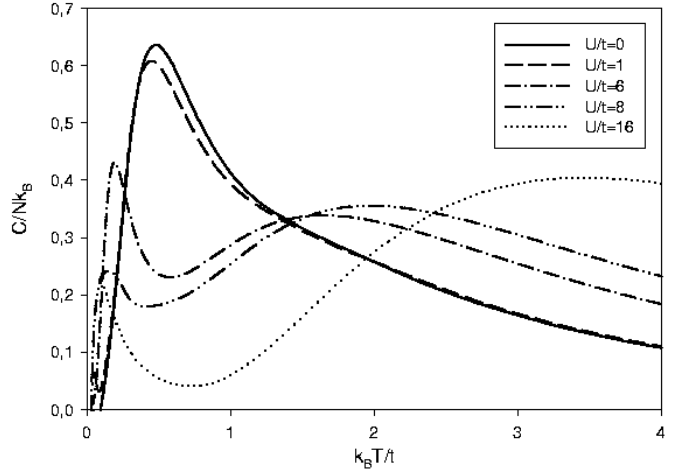,width=90mm,angle=0}
\caption{Specific heat $C/Nk_{B}$ as a function of temperature with $U/t=0$,$0.5$, $4$
and $8$ for the AN of 7 sites.}
\end{figure}

The temperature dependence of the specific heat is shown in Fig. 5 for typical values
of $U/t$. For small values of $U/t$ there is a peak in the specific heat.
Increasing $U/t$, the peak splits into two,
which reflects a rearrangement of the fermionic structure in the system.
The low-temperature peak arises due to low-lying collective excitations, while the
high-temperature broad peak comes from single-particle excitations. 
This behavior is quite general and has been noticed for different
structures \cite{cal,meuDD}.

\begin{figure}
\psfig{figure=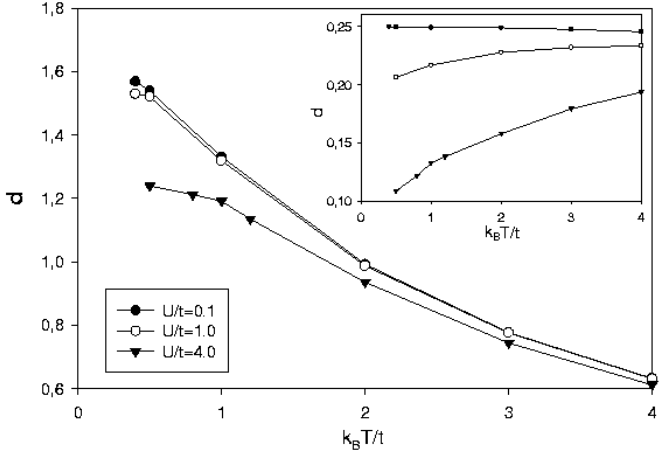,width=90mm,angle=0}
\caption{One-electron transfer $j$ as a function of temperature with $U/t=0.1$, 
$1$ and $4$ for the AN of 124 sites. Inset: Double 
occupancy $d$ as a function of temperature.}
\end{figure}

Next, we study the thermal average of the double occupancy 
$d= \frac{1}{N} \sum_{i} n_{i\uparrow}n_{i\downarrow}$ and of the one-electron
transfer defined by 
\begin{equation}
j = \sum_{<ij> \alpha} c_{i\alpha}^{\dag}c_{j\alpha} = \frac{Ud-\cal H }{t}.
\end{equation}
Fig. 6 shows the temperature dependence of $j$ and $d$ (inset) with $U/t=0.1$, $1$ and $4$ for
the AN with 124 sites. Clearly these functions are related to the
local moment. It is easy to show that for the half-filled band $d=(1-4L_0)/2$.  
If $U/t$ increases, the local moment increases (see Fig. 3a) and the double
occupation and one-electron transfer decrease. In the strong coupling limit 
($U/t >> 1$) each site is occupied by just one electron and the electronic 
itinerancy vanishes. This region is characterized by localized magnetism
and we can apply the Heisenberg model.

\section{Localized magnetism: renormalization group approach}
The fractal and self-similar nature of scale-invariant networks,
generated through a recursive procedure, offer an appropriated way to get rigorous
result at the phase transition. 
The difficulty with the spin-decimation process which leads to the 
renormalization-group equation as found on Bravais lattices due to parameter
proliferation, disappears on scale-invariant networks \cite{RG1,RG2}.  
Due to the fact that fractal lattices are characterized by dilation invariance
but do not have translational invariance, the study on these networks
can with restriction be used to model Bravais lattices.
Particularly self-similar networks can mimick Bravais
lattices for some magnetic models, providing equal critical temperatures and
exponents \cite{RG2,RG3}.

\begin{figure}
\psfig{figure=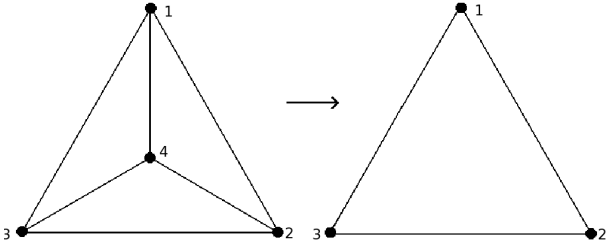,width=35mm,angle=90}
\caption{Renormalization group transformation associated to the AN.}
\end{figure}

The real-space renormalization group (RG) approach has been
applied with success to the study of the anisotropic 
spin $\frac{1}{2}$ Heisenberg model on self-dual
hierarchical lattices \cite{car,mar,meu}. 
Here, we apply this analysis on AN. 
It is defined by the cluster transformation process, 
preserving the Hamiltonian form, as illustrated in Fig. 7.

The dimensionless Hamiltonian is defined by
\begin{equation}
 {\it h} = -\beta {\cal H} = \frac{4J}{k_{B}T} \sum_{<ij>} [(1-\Delta )
(S_{i}^{x}S_{j}^{x} +
S_{i}^{y}S_{j}^{y}) +  S_{i}^{z}S_{j}^{z}] ,
\end{equation}
where $\beta~\equiv~1/k_{B}T$, $\langle ij \rangle $ denotes first-neighboring lattice
sites, $\Delta$ is the anisotropy parameter and the 
$S^{\alpha}_{i} \{ \alpha=x,y,z\}$ is the $\alpha$-spin 1/2 operator 
on site $i$. The RG recurrence equation is obtained by  imposing
\begin{equation}
\exp ({\it h}_{123}^{'} + C) = \trazo \exp ({\it h}_{1234})
\end{equation}
where ${\it h}_{1234}$ and ${\it h}_{123}^{'}$ are, respectively, 
the Hamiltonian of the four-site cluster and of the renormalized three-site cluster,
shown in Fig. 7.
To make the RG equation possible, an additive constant $C$ has been included.
The RG Eq. 6 establishes the relation between the set of parameters ($J,\Delta$)
and the set of renormalized parameters ($J^{'},\Delta^{'}$).
The non-commutativity between the Hamiltonians associated with neighboring clusters
is neglect, and therefore our results are approximations for all temperatures,
only being asymptotically exact at high temperatures \cite{mar}. 

\begin{figure} 
\psfig{figure=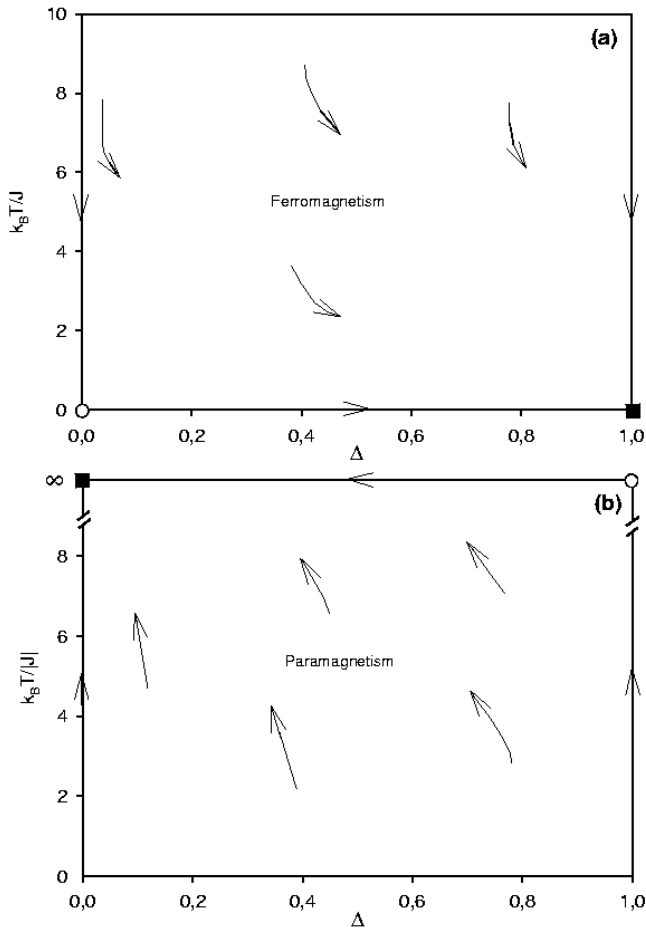,width=90mm,angle=0}
\caption{Flow diagram for the Apollonian cell of Fig. 7. Open circles,
and full squares respectively, denote,  the semi-stable and fully stable fixed points.
(a) ferromagnetic case ($J~>~0$). (b) antiferromagnetic ($J~<~0$) case.}
\end{figure}

Defining $K~\equiv~J/k_{B}T$ and expanding $exp({\it h}_{123}^{'})$ as 
\begin{eqnarray}
\exp ({\it h}_{123}^{'} + C) \! = \! a^{'}+4 b^{'}(S_{1}^{x}S_{2}^{x} +
S_{1}^{y}S_{2}^{y}+S_{1}^{x}S_{3}^{x} + S_{1}^{y}S_{3}^{y} \nonumber \\
 +S_{2}^{x}S_{3}^{x} + S_{2}^{y}S_{3}^{y}) 
+ 4 c^{'} (S_{1}^{z}S_{2}^{z}+S_{1}^{z}S_{3}^{z}+S_{2}^{z}S_{3}^{z}),
\end{eqnarray}
we obtain that
\begin{eqnarray}
4K^{'}&=& \ln ( \frac{a^{'}+3c^{'}}{a^{'}-c^{'}-2b^{'}}) + \frac{1}{3}
\ln ( \frac{c^{'}-a^{'}+2b^{'}}{c^{'}-a^{'}-4b^{'}} ) \nonumber \\
6K^{'}\Delta^{'}&=& \ln ( \frac{c^{'}-a^{'}-4b^{'}}{c^{'}-a^{'}+2b^{'}} ) \nonumber \\
C&=&-3k^{'}+ \ln (a^{'}+3c^{'}). 
\end{eqnarray}
  Analogously,
\begin{eqnarray}
\exp ({\it h}_{1234}) \! =&a + 4 \sum_{(i<j)} [b(S_{i}^{x}S_{j}^{x} + S_{i}^{y}S_{j}^{y})
 + c S_{i}^{z}S_{j}^{z}]  \nonumber \\
&+16 \sum_{(i<j)\neq (k<l)}[ d(S_{i}^{x}S_{j}^{x} + S_{i}^{y}S_{j}^{y})S_{k}^{z}S_{l}^{z} \nonumber \\ 
& +g (S_{i}^{x}S_{j}^{x} + S_{i}^{y}S_{j}^{y})(S_{k}^{x}S_{l}^{x} +
S_{k}^{y}S_{l}^{y}) ]  \nonumber  \\
&+ 16~f~S_{1}^{z}S_{2}^{z}S_{3}^{z}S_{4}^{z}
\end{eqnarray}
where $a,b,c,d,g$ and $f$ are functions of $K$ and $\Delta$. 
It is easy to see from Eq. 6 that $a^{'}=2~a$, $b^{'}=2~b$ and $c^{'}=2~c$.
The set of parameters ($K^{'},\Delta^{'}$)
as functions of ($K,\Delta$) can be determined by diagonalizing ${\it h}_{123}$ and 
${\it h}_{1234}$ and using Eqs. 6-9. After some calculus we obtain the analytical RG 
equation as
\begin{eqnarray}
a^{'}\! &=&\! \frac{1}{4} [\! e^{6K}\! +\! e^{6K\Delta }\! +\! 3e^{-2K\Delta }\! +\!
\frac{e^{-2K}}{2}\! (\! 3+e^{8K\Delta }+2e^{-4K\Delta }\! )\! ]  \nonumber \\
b^{'}\! &=&\! \frac{1}{4} [ \frac{e^{6K\Delta }-e^{-2K\Delta }}{2}+
\frac{e^{-2K}}{3}( e^{8K\Delta }-e^{-4K\Delta }) ] \nonumber \\
c^{'}\! &=&\! \frac{1}{4} [ e^{6K}- \frac{e^{-2K}}{6}( 3+e^{8K\Delta }+2e^{-4K\Delta })] 
\end{eqnarray}
where $K^{'}, \Delta^{'}$ are functions of $c^{'}$ $a^{'}, b^{'}$ given by Eq. 8. 
We observe that the Ising and isotropic Heisenberg models are mapped into
themselves. The RG recurrence is simplified as
\begin{equation}
k^{'}= \frac{1}{4} \ln ( \frac{1+e^{6K}}{1+e^{-2K}} )
\end{equation}
and
\begin{equation}
k^{'}= \frac{1}{6} \ln ( \frac{3+5e^{8K}}{6+2e^{-4K}} )
\end{equation}
corresponding to the Ising ($\Delta =1$) and the isotropic Heisenberg ($\Delta =0$)
limits, respectively.

Fig. 8 shows the phase diagram in the ($k_{B}T/J,\Delta$) space. 
In the ferromagnetic case ($k_{B}T/J \! > \! 0$), we always observe 
the existence of ferromagnetism(F) independent of $\Delta$. This result has been
found on AN for the Ising model using the transfer matrix method \cite{hans2} and 
on other scale-free lattices \cite{alek}. 
Those results agree very well with ours. Here, we observe that 
the quantum fluctuations of the $XY$ part of our Hamiltonian
does not destroy this ordering generated by the topology of the AN.
In the antiferromagnetic case ($k_{B}T/J\! <\! 0$), we verify the
non-existence of the ordered phase for all $\Delta$ at finite $T$. 
Again, the results obtained are similar and consistent with the 
Ising limit \cite{hans2}.

\section{Conclusions}
In conclusion, we have analyzed strongly correlated electron systems on AN.
The ground state and thermodynamic properties of the Hubbard model have been
studied using exact diagonalization calculations and quantum Monte Carlo.
The dependence on the ratio $U/t$ of the specific heat, the magnetic susceptibility,
the spin-spin correlation function, the double occupancy and the one-electron transfer
support several types of magnetic behavior. 
We have also studied the magnetic properties of the anisotropic
spin $\frac{1}{2}$ Heisenberg model on the AN using the real-space RG approach.
For ferromagnetic coupling, we always observe the existence of 
ferromagnetism independent on temperature and anisotropy parameter $\Delta$. 
As opposed to other structures \cite{mar,meu}, the topology of the AN favors
the ferromagnetic order and the quantum fluctuations do not destroy this ordering. 
For antiferromagnetic coupling, we find a paramagnetic phase for all
$\Delta$ at finite $T$. A similar result has been found on AN for the Ising model
using the transfer matrix method \cite{hans2}.

\section*{Acknowledgments}

Authors thank P. G. Lind and E. Parteli for their generous help.
This work was supported by CNPq (Brazil), DAAD (Germany) and the Max Planck prize.

\end{document}